# AES CCMP Algorithm with N-Way Interleaved Cipher Block Chaining

**Zadia Codabux-Rossan ***
*Faculty of Engineering,*
*University of Mauritius*
Email: *z.codabux@uom.ac.mu*

**M. Razvi Doomun**
*Faculty of Engineering,*
*University of Mauritius*
Email: *r.doomun@uom.ac.mu*

**Abstract**

Nowadays, the increased use of battery-powered mobile appliances and the urge to access time-sensitive data anytime anywhere has fuelled a high demand for wireless networks. However, wireless networks are susceptible to intrusion and security problems. There is an inherent need to secure the wireless data communication to ensure the confidentiality, authenticity, integrity and non repudiation of the data being exchanged. On the other hand, the computation and the resultant energy consumption to achieve sufficient security can be high. Encryption algorithms are generally computationally intensive, and consume a significant amount of computing resources (such as CPU time, memory, and battery power). Considering the limited resources on wireless devices, it is crucial that security protocols be implemented efficiently.

This manuscript focuses on how energy consumption is impacted by the use of unoptimised AES-CCMP algorithms and proposes an optimized AES CCMP algorithm using 2-way interleaving that does not compromise the security of wireless communication sessions. There is also analysis of the performance of AES (a.k.a. Rijndael) in its AES–CCMP implementation. The 2-way interleaving technique is an optimization of the CBC-MAC that is investigated using two performance metrics (namely encryption time and throughput).

**Keywords**: IEEE 802.11i security, AES-CCMP, Optimization, Interleaved Cipher Block Chaining

**For correspondences and reprints*

# I. INTRODUCTION

With the maturation of Industry standards and the deployment of lightweight wireless hardware across a broad market section, wireless technology has come of age. Wireless networks allow computers and peripherals to communicate using radio RF transmissions as an alternative to conventional network cabling. Wireless networks use electromagnetic energy travelling in space as an information transport mechanism between devices and the traditional wired network infrastructure, or among wireless devices (when in communicating in an ad-hoc mode). Advances in wireless network technologies are frequent, which is made evident by the vast number of publications in the field of IEEE 802.11 Wireless Local Area Networks (WLANs), mobile ad-hoc networks and wireless sensor networks. Deployment of wireless networks is on the rise. Its popularity is due to the ability to provide communications rapidly, further augmented by features like ubiquity and mobility.

The rapid growth and integration of these systems into a wide range of networks architectures, and for a wide variety of applications, drives the need for wireless security approaches to support the requirements of a wide variety of customers. Due to the broadcast nature of wireless radio signals, wireless networks are implicitly vulnerable to several types of network attacks. Anyone within range of a wireless device's transmissions is able to passively listen to (or eavesdrop on) the signals, and could potentially access the information contained in the signals. It is also possible to actively transmit signals that can attack the network. Wireless networks are therefore vulnerable to many kinds of unique security threats (not present in conventional wireless networks), and require strong countermeasures to overcome these threats.

It is imperative to provide adequate security services to wireless networks. But, providing security for wireless devices is challenging, because wireless devices have limited resources such as low CPU speeds, limited memory capacity, and most importantly limited battery power. Making efficient use of battery power is a multifaceted research topic by itself, but designing efficient security services that make conservative use of battery-powered devices is a real challenge. Security services rely on cryptographic and mathematical functions that are known to be computationally intensive. To implement security mechanisms for such resource-limited and battery-powered devices, innovative techniques are required to find the best trade-off between optimizing security strength to thwart existing security attacks and conserving maximum battery power to expand the operational lifetime of these devices.

The structure of the paper is organised as follows: AES and AES-CCMP is explained in section II; a review of related works is given in section III; the interleaved CBC technique is proposed in section IV; the experimental set and results & discussions follow in section V and VI, respectively; the work is concluded in section VII.

# II. AES OVERVIEW

## A. AES

AES is a symmetric iterated block cipher, meaning that the same key is used for both encryption and decryption, multiple passes are made over the data for encryption, and the clear text is encrypted in discrete fixed length blocks. In FIPS Publication 197 [FIPS 197, 2001], the U.S. NIST officially endorsed the Rijndael algorithm to be used as the AES in cryptographic systems for the protection of unclassified, non-public information throughout Federal Agencies. The AES CCMP implementation uses an AES 128-bit key and 128-bit block size. Per the FIPS 197 standard [FIPS 197, 2001], the AES algorithm (a block cipher) uses blocks of 128 bits; cipher keys with lengths of 128, 192 and 256 bits; as well as a number of rounds 10, 12 and 14 respectively. The security of the AES algorithm depends on the number of "Rijndael" rounds (rounds are a measure of the number of repetitions through an encryption algorithm that information goes through for transformation from plain-text to cipher-text). As more rounds are involved in the overall transformation, the typically result is a higher encryption strength [W Roche, 2006].

Each Rijndael round is composed of four operations: byte substitution, shift rows, mix columns, and add round key (with some exceptions in the final round, because the final round does not include the mix columns operation). Also before the first round, an "add round key" is required, which could be considered overhead for each encryption task (or for each data packet when encryption is used for communications). "Byte substitution" is an invertible, non linear transformation. It uses 16 identical S-Boxes for independently mapping each byte of the state into another byte. "Shift rows" operates on the rows of the state, rotating the second, third and fourth row of the state by one, two, or three bytes respectively. "Mix columns" performs a modular polynomial multiplication in GF $(2^8)$ on each column, and it is a resource intensive transformation. "Add round key" performs an XOR with each state and round key. The round key generation (also known as key expansion) includes S-Box substitution, word rotations and XOR operations performed on the encryption key, before encryption starts.

## B. AES CCMP

The IEEE Std. 802.11-2007 Amendment 6 (formerly IEEE 802.11i-2004) defines an encryption method based on the AES. AES-based encryption can be used in a number of different modes or algorithms. The AES mode that has been chosen for 802.11 is AES-CCM. The AES-CCM Protocol is a data-confidentiality protocol that provides both packet authentication and encryption. For confidentiality, AES-CCMP uses AES in counter mode. For authentication and integrity, AES-CCMP uses CBC-MAC. In IEEE Std. 802.11-2007 Amendment 6, AES-CCMP uses a 128-bit key. AES-CCMP protects some fields that aren't encrypted. The additional parts of the IEEE 802.11 frame that get protected are known as AAD. AAD includes the packets source and destination addresses. AES-CCMP protects AAD against replay attacks (thereby preventing an attacker from retransmit a message that was encrypted with AES-CCMP and having that packet be accepted as a valid packet by the receiver). AES-CCM is intended for use in a packet environment (i.e., when all of the data is available in memory and can be broken into discrete packets of information before the AES-CCM protocol is applied); AES-CCM is not designed to support partial processing or stream processing. The input into the AES-CCM protocol includes three elements:

1) data that will be both authenticated and encrypted, called the payload
2) associated data (e.g., a header, that will be authenticated but not encrypted)
3) A unique value, called a nonce, that is assigned to the payload and the associated data [M Dworkin, 2004].

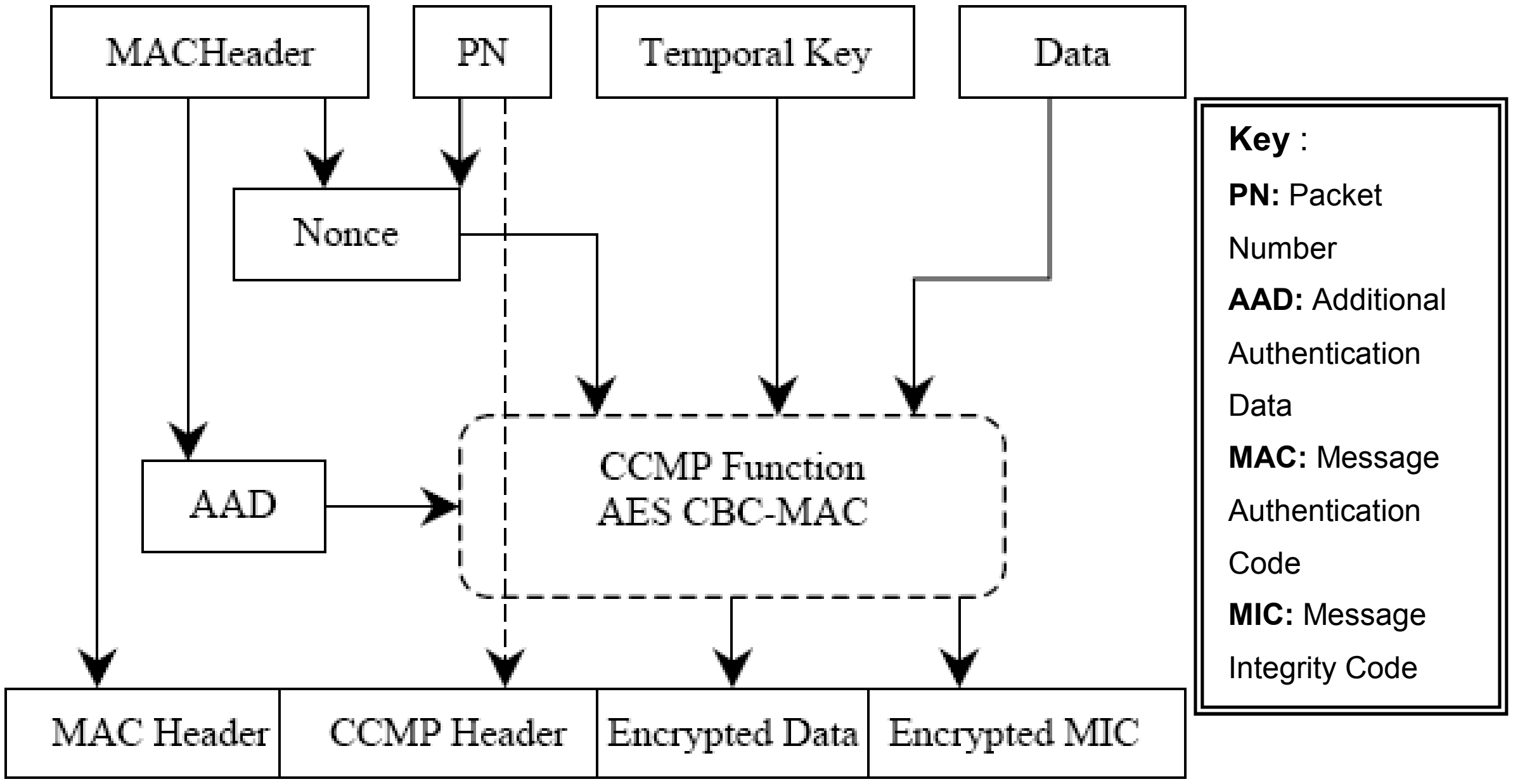

*Fig I: AES-CCMP [R Doomun et al., 2007]*

## C. AES-CCMP in Wireless Network Security

Wireless devices have limited resources such as low-speed CPUs, limited available memory, and most importantly limited battery power. The pace of advancements in battery technologies has not kept up with that of wireless technologies. This implies that mobile devices typically operate on a frugal power budget and therefore computationally intensive encryption/decryption algorithms and the related security parameters may not be supported. It must also be noted that there are several applications such as wireless sensor networks where the battery power limitation is extreme and re-charging or changing out of drained batteries may be near impossible. Therefore, the primary challenge in providing security in low power mobile wireless devices lies in the conflicting interest between minimizing power consumption and maximizing security. From previous works, there are a number of algorithms for wireless networks.

It can be safely assumed that by doing more computations, one can achieve a higher amount of security. For example, the strength of an encryption schemes depend on the size of the key and the number of encryption rounds [J. Nechvatal et Al., 1999]. Larger key sizes/rounds produce higher levels of security at the cost of additional power consumption. To illustrate this point, Figure II shows the trade-off between vulnerability and power consumption by varying the number of encryption rounds.

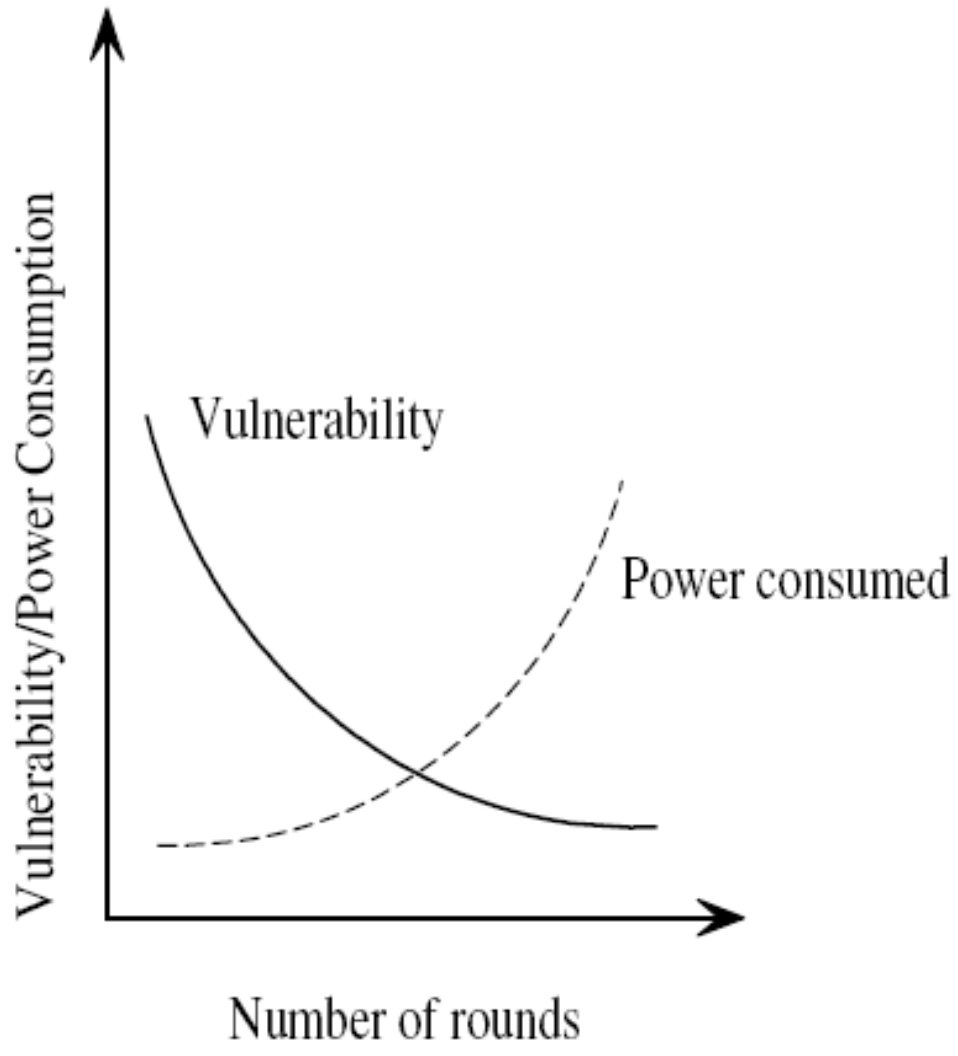

*Fig II: Security vs. battery power consumption trade-off [R Chandramouli et al., 2005]*

Due to the efficiency and performance of Rijndael, AES-CCMP is a good candidate for wireless network devices. Additionally, there have been numerous efforts previously developed as hardware and software implementations of AES-CCMP that have been carried out in order to optimize the AES-CCMP encryption algorithm to reduce power consumption.

## III. Related Work

In this chapter, a review of recent efforts concerning energy consumption of cryptographic mechanisms is presented and the research conducted is critically examined.

The main sources of energy consumption during a secure wireless transaction are:

(i) cryptographic computations used to establish secure sessions and for encryption and authentication
(ii) cryptographic computations used for performing secure data transactions [R Karri et al., 2003].

Security in wireless networks can be achieved by security protocols at different levels of the protocol stack for example WEP at the Link Layer, IPSec at the Network Layer, TLS/SSL and WTLS at the Transport Layer and so on. Security protocols are made up of cryptographic algorithms which can be categorized as asymmetric and symmetric algorithms (used for authentication and privacy purposes) and hash algorithms (used for message integrity) [N Potlapally et al., 2003]. One of the main challenges of mobile wireless systems is the mismatch between security requirements and available battery capabilities. Mobile devices are continuously decreasing in size; this is coupled with an increase in the demand for more security (especially for wireless transactions) and efficient battery energy management schemes.

A number of research publications have focused on analyzing the energy consumption of different encryption algorithms rather than finding ways to optimize battery life.

There are some works which have compared the energy consumption of AES/Rijndael with other algorithms. [P Prasithsangaree et al., 2003] compared AES and RC4 and proved that AES is more suitable for devices with low processing power, such as wireless devices. Performance and energy consumption of the following block ciphers Rijndael (AES), RC6, Serpent, Twofish and XTEA were evaluated by [J Groβschädl et al., 2007] with an emphasis on lightweight software implementations. [C.T. Hager et al., 2005] describes a study that compared encryption algorithms namely RC2, BLOWFISH, XTEA and AES to investigate the performance in terms of latency and throughput and energy consumption of block ciphers on a resource limited handheld device, the PDA.

Research has also been carried out with protocols and algorithms other than AES, to investigate their energy-efficiency and propose ways to optimize them to minimize their energy consumption. [P Ni et al., 2004] reviewed the energy consumption of IPSec. The papers by [N Potlapally et al., 2003] and [N Potlapally et al., 2006] compared the energy consumption among the three types of cryptographic algorithms namely symmetric, asymmetric and hash algorithms. [R Karri et al., 2003] conducted a study for the energy cost of session negotiation protocols used by IPSec and WTLS protocols and proposed techniques to optimize the energy consumption during the session negotiation. [P Agrawal, 1998] conducted a study of the energy savings at various levels of the TCP/IP protocol stack for wireless systems.

There are also numerous software strategies to optimize energy consumption in wireless networks that have been proposed by different academic and research organizations. According to [P Prasithsangaree et al., 2003], a common way to minimize wasted transmission energy in a communication protocol is to send a short probe first to determine if conditions for data transfer is optimum and then send data. The probes are encrypted with an encryption algorithm which doesn't consume much energy. A proposal by [C N Mathur et al., 2006] consisted of proposing a novel block cipher, HD Cipher to replace AES in the CCMP. HD Cipher has a 288-bit keystream and therefore has fewer encryptions per frame. Energy savings realized by HD Cipher were of the order of 40% over the use of AES. Computation offloading on a handheld in a wireless LAN secured by IPSec was investigated by [Z Li et al., 2002]. [N Potlapally et al., 2003] and [R Doomun et al., 2007] both studied the use of an adaptive resource-aware security protocol which altered its behavior based on the operating environment. [N Potlapally et al., 2006] mentioned that there are 2 software techniques for improving performance, namely table look-ups and loop unrolling. The NOVSF code hopping technique was proposed by [H Cam et al., 2002]. NOVSF takes advantage of the time slots and assigns data blocks to different time slots in every session and therefore increases communication security without additional energy.

In addition to the software approaches proposed earlier, there exists many hardware schemes to optimize power-security efficiency. [A Aziz et al., 2007] implements a fast, efficient, low-power FPGA for AES-CCM whereby the computational intensive cryptographic processes are offloaded from the main processor. [K Atasu et al., 2004] proposed an implementation which uses the ARM core architecture and a new implementation for the new MixColumn implementation. [C Mucci et al., 2007] demonstrates the implementation of the AES/Rijndael algorithm on the DReAM architecture which is a dynamically reconfigurable architecture. [P Hämäläinen et al., 2006] presents the design and implementation of a compact 8-bit AES ASIC encryption core suitable for low-cost and low-power devices.

## IV. Interleaved Cipher Block Chaining

Interleaved encryption is the processing of the encryption of a message as multiple independent messages block of known size, with *N* different IVs, generally treating every nth block as part of a single message. CBC is difficult to parallelize, which led to the development of Interleaved CBC (ICBC), in which multiple streams of CBC encryption are interleaved [K Gaj et al., 2000]. The encryption of the next block of data can start as soon as the block *N* positions earlier have been encrypted.

For this thesis, we are proposing the implementation of two-way CBC interleaving as an optimisation in the AES CCMP encryption. In two-way interleaved chaining, the first, third and every two block thereafter is encrypted in CBC mode. The second, fourth and every two block thereafter is encrypted as another stream. An example of such mode is the interleaved CBC mode shown in Figure III.

The benefits of interleaved modes, such as the interleaved CBC mode is that they have a potential to offer security of feedback modes combined with the performance of non-feedback modes. Interleaving allows delivering of high performance mainly in terms of gain in speed while maintaining level of security. However, as shown in Figure III, for the ICBC, 2 IVs are required to be transmitted to the receiver. Moreover, more than 1 result is obtained as the outcome of ICBC. An additional computation is required to merge the multiple results into 1. The two-way interleaved chaining will produce 2 results at the end. The multiple results will be XORed to produce a single MIC as the outcome.

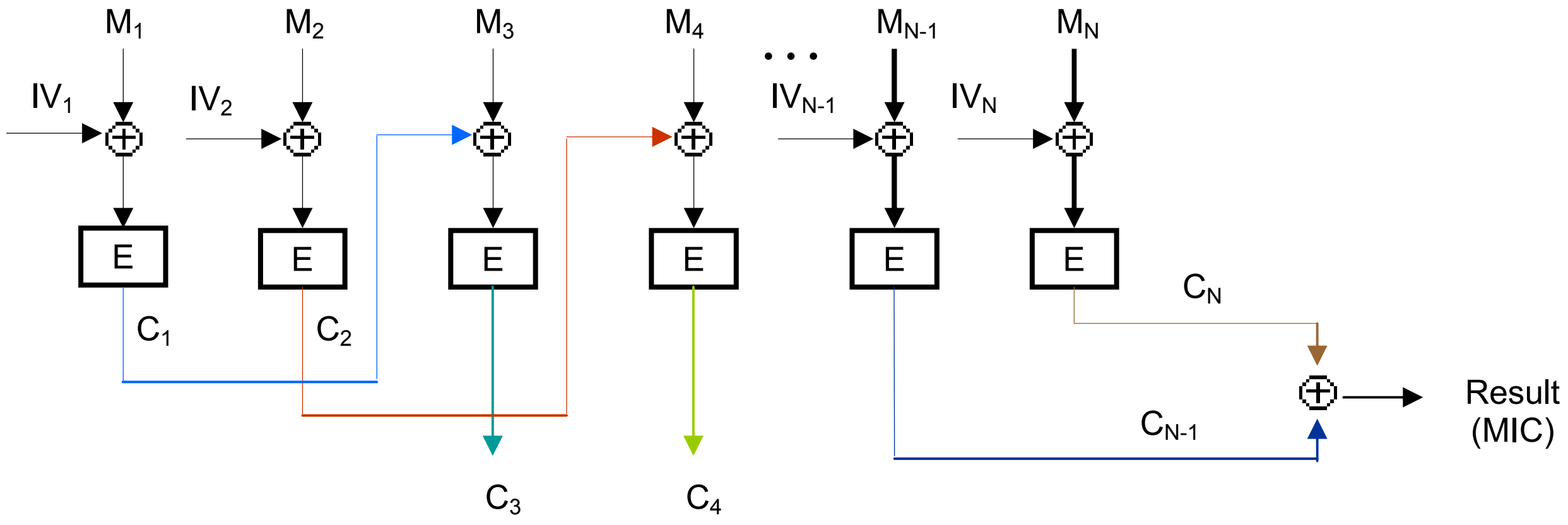

*Fig III: Interleaved CBC Mode*

Such modes are likely to be considered by NIST for standardization as future AES operating modes and become a part of other standardization efforts.

## V. Experimental Setup

The tests have been done on an HP Compaq Presario V2000, which is equipped with an Intel Pentium Mobile Processor Centrino 1.6 GHz CPU working at a (constant) clock rate of 598.5 MHz and a physical RAM of 512 MB. The operating system was Microsoft Windows XP Professional Version 2002. During the simulations, there were no other tasks running on the system except the system tasks.

The coding of the program was performed in the C++ language, which is a high-level language defined at higher abstract levels and is programmer-friendly. C++ was compiled using DEV C++ 4.9.9.2. High-level synthesis helped in realizing the project objectives in a lesser amount of time. The main advantage of using a high-level language is the code portability.

For obtaining the energy consumption of the encryption algorithm it was necessary to determine several factors, which can be used for a general comparison. In our experiment, the performance metrics measured are encryption time and throughput.

The encryption time is the time that an encryption algorithm takes to produce a ciphertext from a plaintext. It is calculated as the total of CBC MAC encryption time and Counter mode encryption time.

Encryption time is used to calculate the throughput of an encryption scheme that indicates the speed of encryption. The throughput of the encryption scheme is calculated as the total plaintext in bytes encrypted divided by the encryption time.

# VI. Results & Discussions

## A. Unoptimized AES CCMP

AES CCMP has been implemented by calculating the MIC tag and then encrypting the plaintext and MIC tag in counter mode to get the cipher text. While increasing plaintext size, the encryption time and throughput are measured to determine the time it takes to encrypt plaintext converting it into cipher text, using the unoptimised version of the AES CCMP software.

- **Encryption Time Measurement**

The encryption time measured is the total time of the CBC MAC time summed with the Counter mode encryption time. The CBC MAC encryption time is the total time of the MIC IV calculation time added with the headers calculation time and the MIC tag calculation time.

**CBC MAC encryption time** = construct_mic_iv() time + construct_mic_header1() time + construct_mic_header2() time + calculate_mic() time

The Counter mode encryption time is the total time of the Counter calculation time and the data and MIC tag encrypted in counter mode time.

**Counter mode encryption time** = construct_ctr_preload() time + encrypt_mpdu() time

For obtaining an average value, it was necessary to run several tests based on the same plaintext to minimize the influence of measurement errors, which were generated by the system tasks, on the result. The results of these measurements were averaged to get the encryption time for 1 plaintext. These tests comprised measuring the time the algorithms needed to encrypt 1, 2, 3 and 4 blocks, each of 16 bytes length. The results are presented in the table below and displayed graphically in Figure IV. With an increasing plaintext data size (in bytes), the encryption time increases. Increasing the plaintext data by 4 times results in increasing the encryption time by 340.7%.

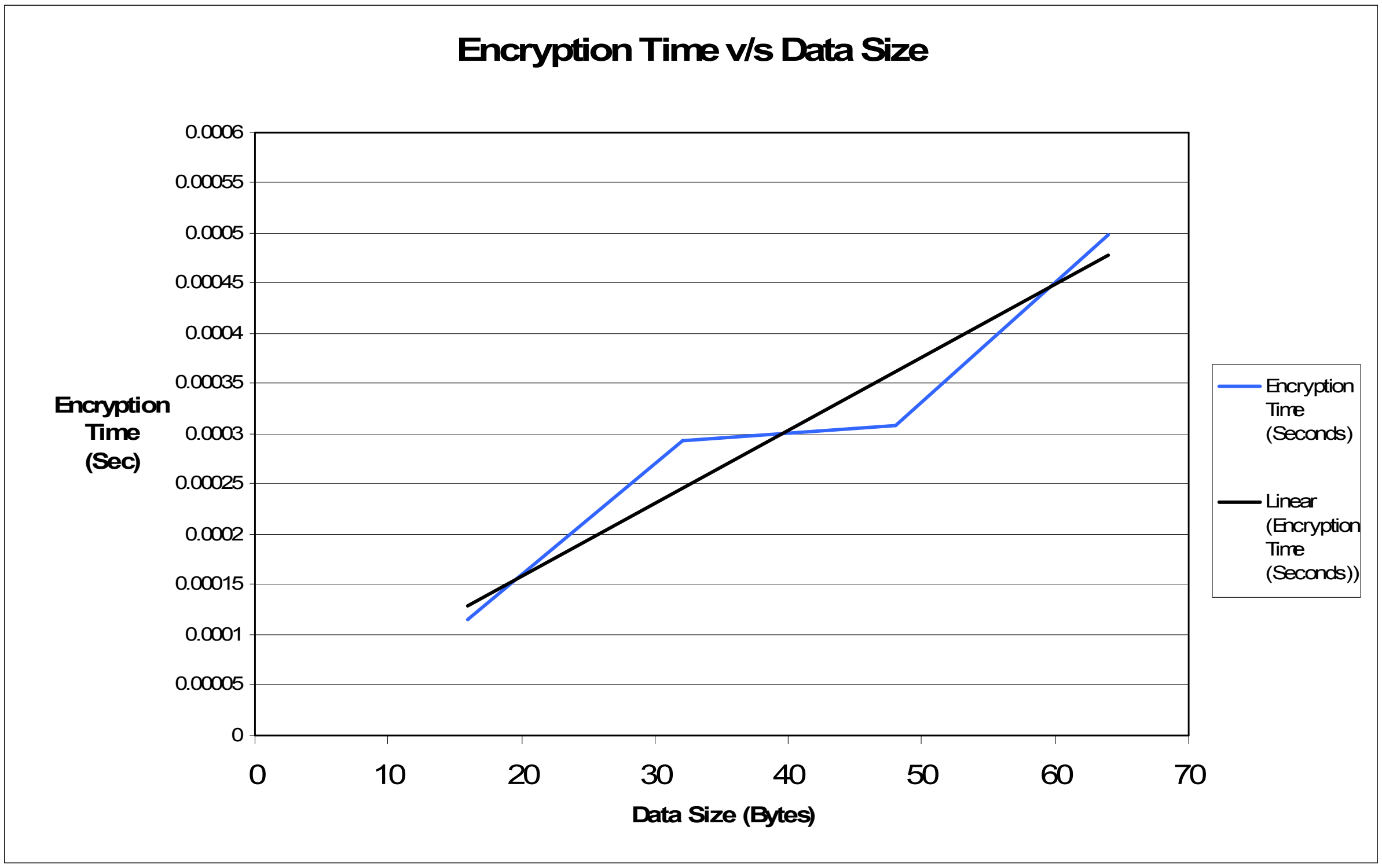

*Fig IV: Encryption Time v/s Data Size for unoptimised AES CCMP*

- **Throughput Measurement**

The throughput was measured by dividing the length of the data size by the encryption time. An increase in the plaintext data size will cause a decrease in the throughput. Increasing the plaintext data by 4 times decreases the encryption time by 331.7%.

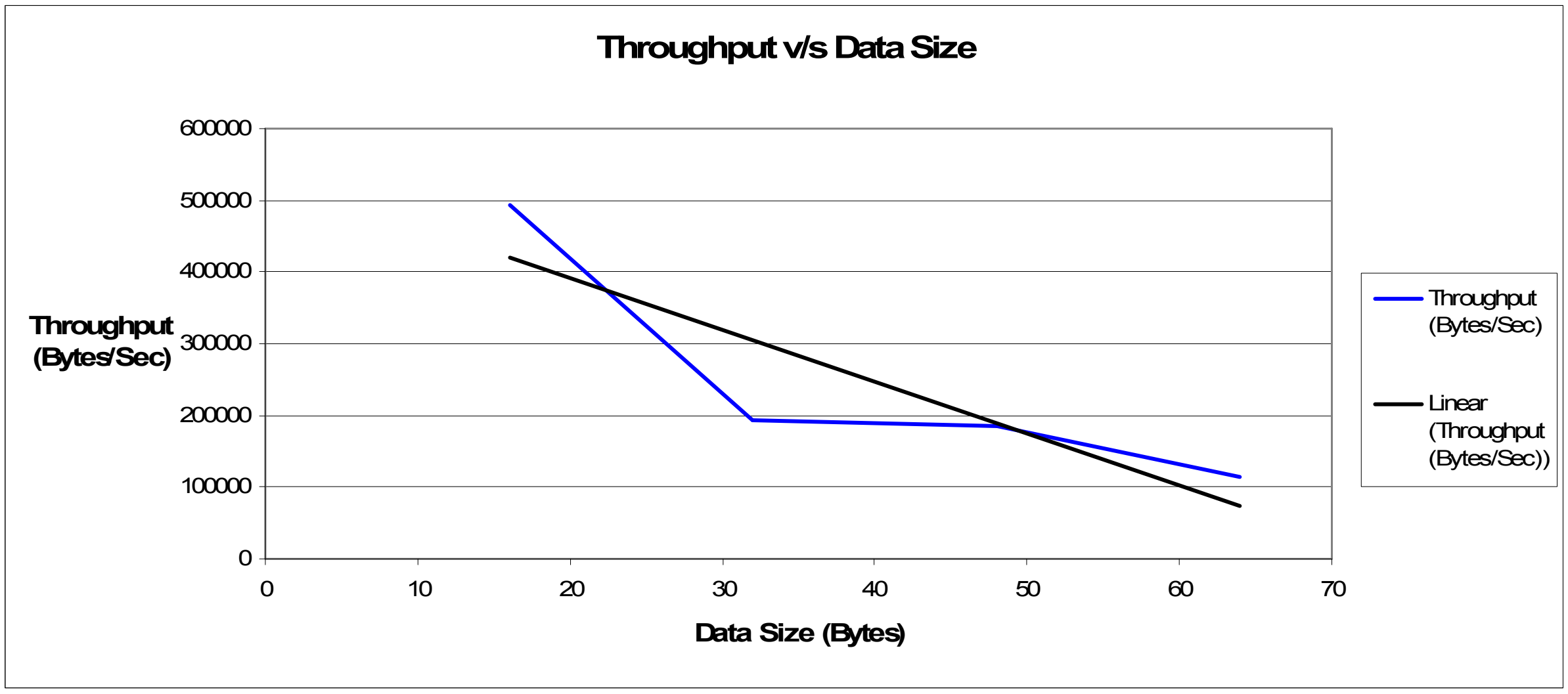

*Fig V: Throughput v/s Data Size for unoptimised AES CCMP*

## B. 2 Way Interleaved ICBC

AES CCMP has been implemented using 2-way interleaving to test for optimization. While increasing the plaintext size, the encryption time and throughput are measured to determine the time it takes to encrypt plaintext converting it into cipher text, using the 2-way interleaved version of the AES CCMP software.

- **Encryption Time Measurement**

It should be noted that the CBC MAC encryption time takes up to 80% of the total encryption time and according to the equation for the CBC MAC encryption time, the IV and headers take up to about 3% and the MIC tag calculation takes up 97% of the total time respectively. With the 2 way ICBC optimization, the encryption time for the IV and headers will remain similar as for the unoptimised AES CCMP, and the Counter mode time also will be same as if no modifications had been made to that part of the design. However, as the encryption of the plaintext will be done in parallel, the MIC tag calculation time in expected to decrease by approximately 50%. Figure VI shows that the general trend is an increase as the data size in bytes increases.

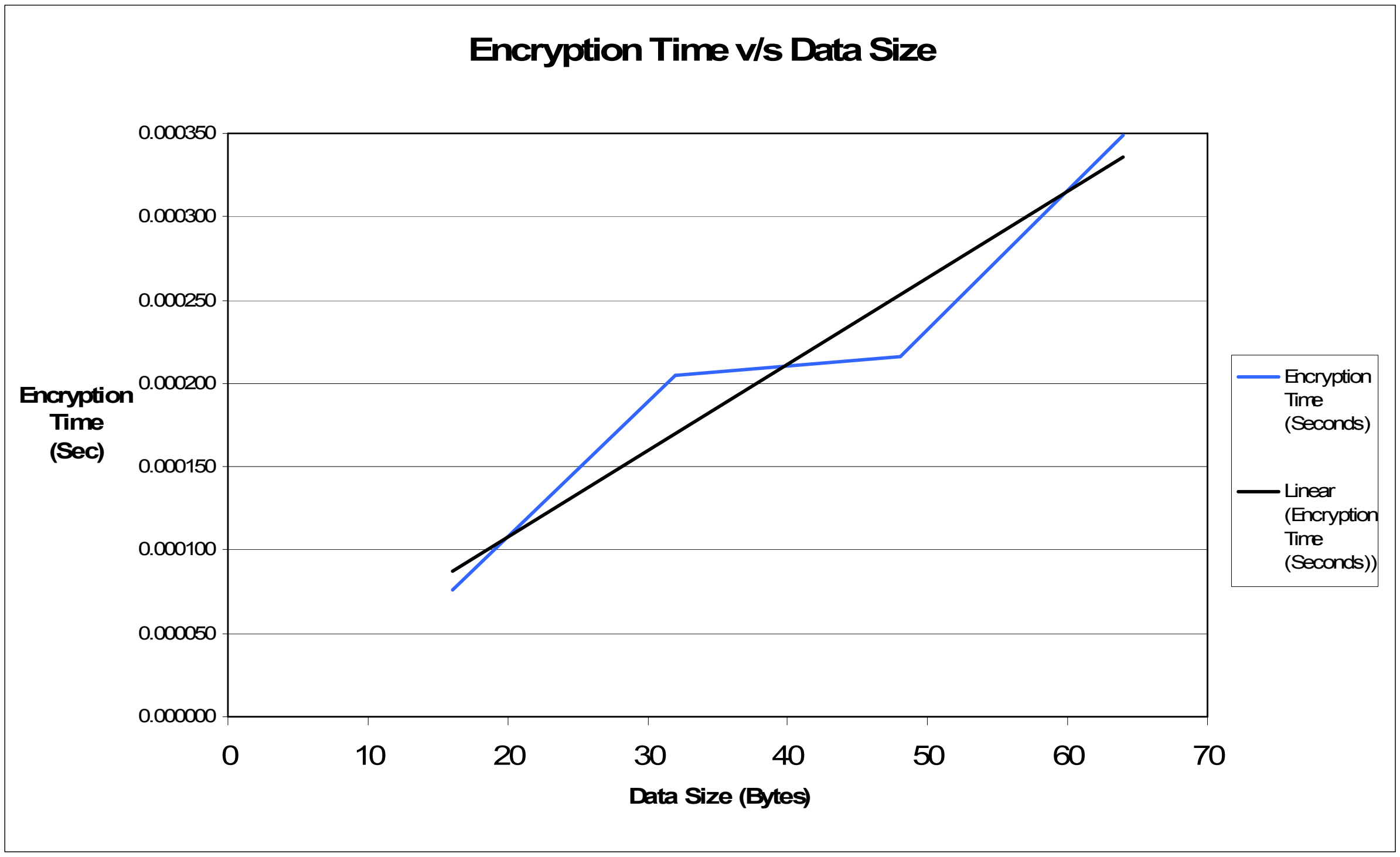

*Fig VI: Encryption Time v/s Data Size for 2 Way Interleaved AES CCMP*

- **Throughput Measurement**

The throughput for the 2 Way Interleaved AES CCMP is as shown in Figure VII:

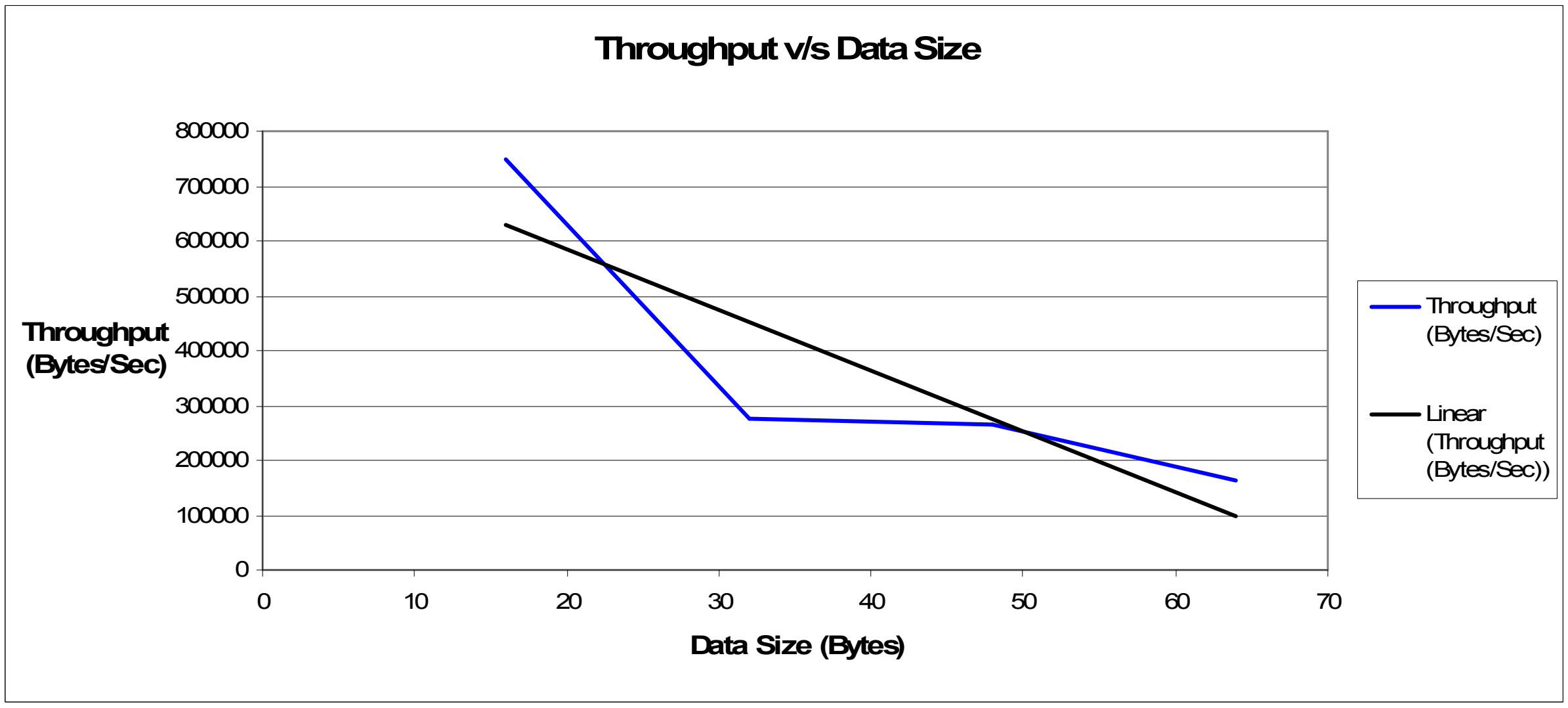

*Fig VII: Throughput v/s Data Size for 2 Way Interleaved AES CCMP*

## C. Unoptimised AES CCMP V/S 2-Way ICBC

- **Encryption Time Measurement**

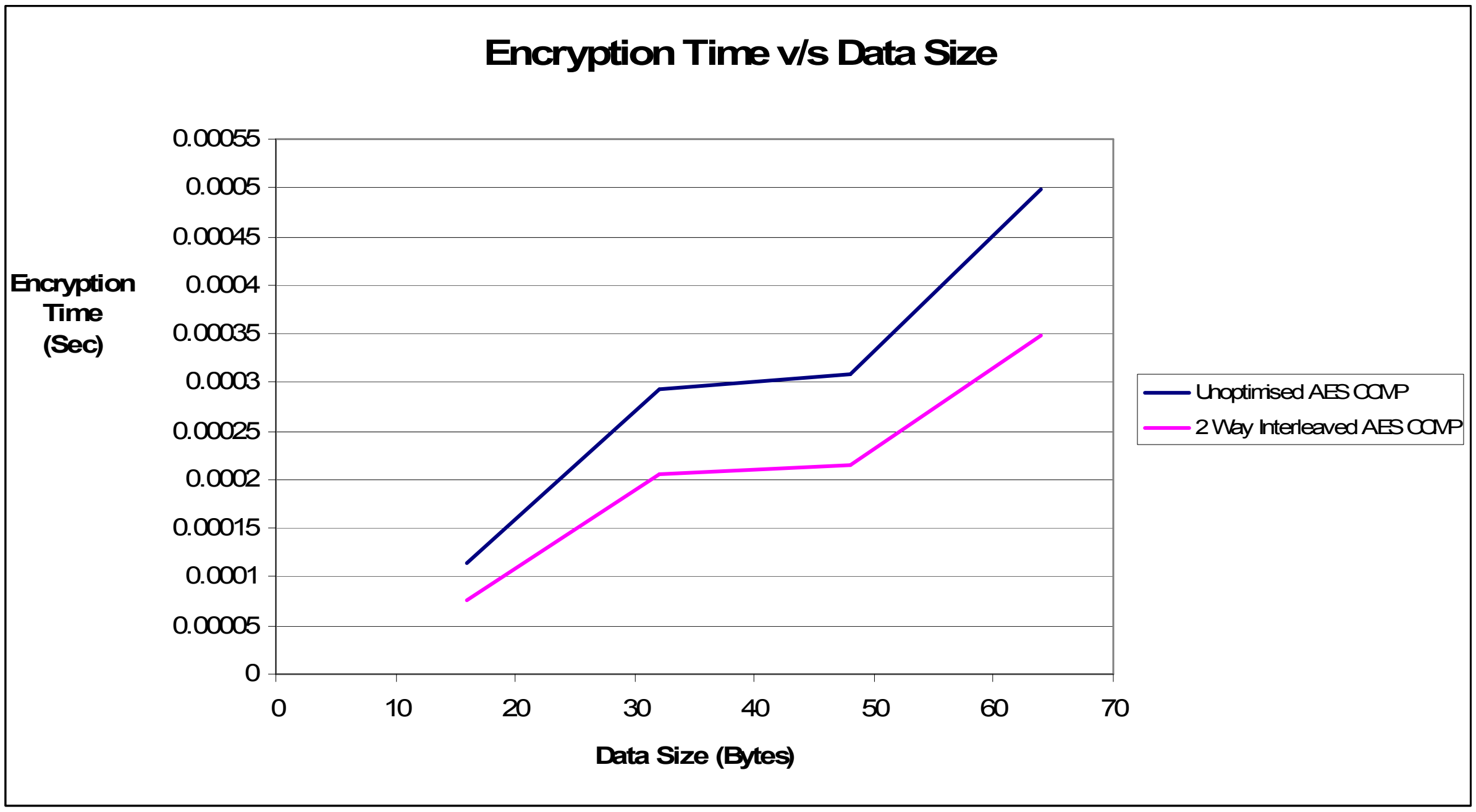

*Fig VIII: Unoptimised AES CCMP v/s 2 Way Interleaved AES CCMP for encryption time metric*

The above graph, Figure VIII, clearly shows that it takes less time to encrypt the same amount of data with the 2 way interleaved CBC algorithms rather than the unoptimised AES CCMP. The approximate percentage decrease in time taken is about 30%.

- **Throughput Measurement**

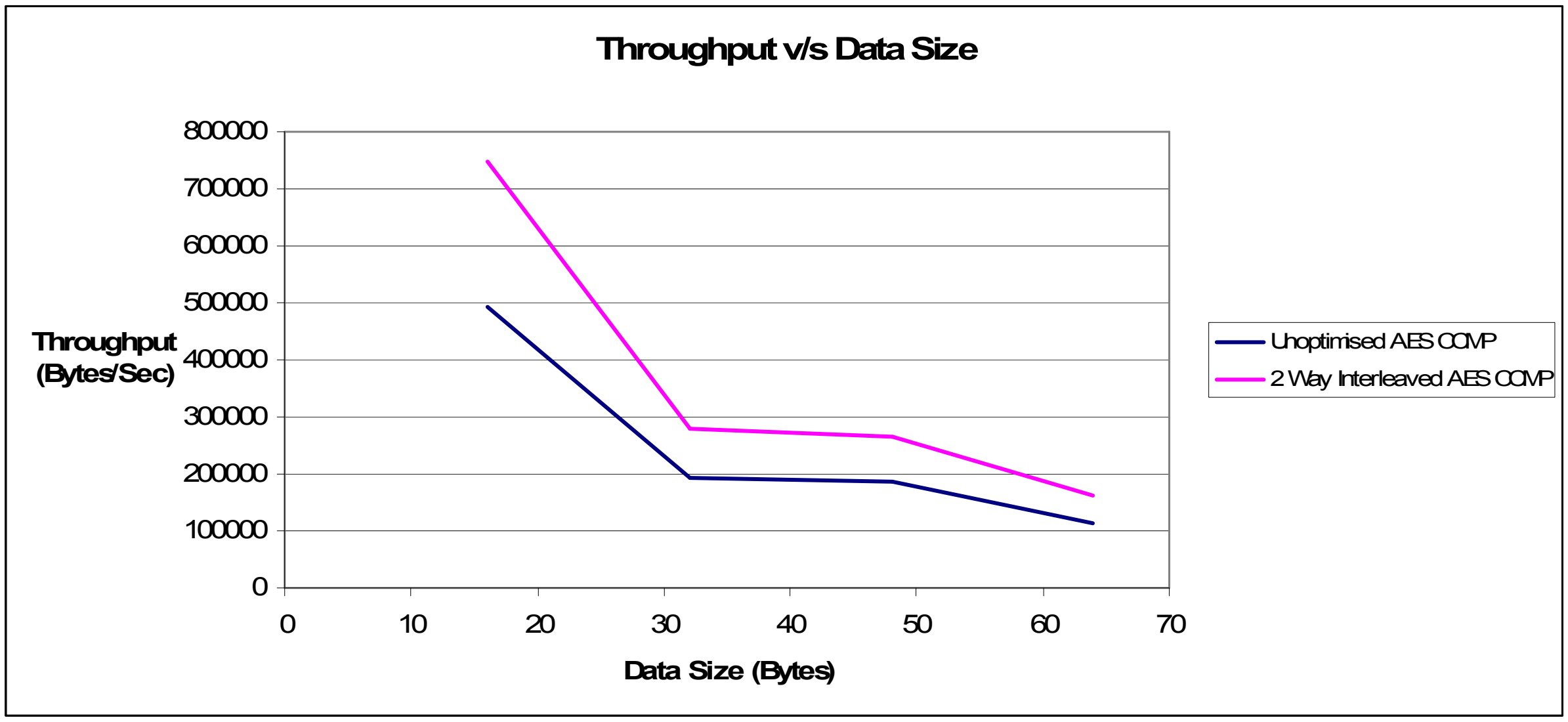

*Fig IX: Unoptimised AES CCMP v/s 2 Way Interleaved AES CCMP for throughput metric*

With increasing data size, the throughput decreases. The 2 way interleaved CBC algorithms encrypt more data per unit time rather than the unoptimised AES CCMP. More data is encrypted for 2 way interleaved CBC algorithm rather than unoptimised AES CCMP during the same lapse of time.

## VII. Conclusion & Future Works

In this paper "unoptimised AES CCMP" was first implemented and performance metrics like encryption time and throughput were evaluated for increasing number of data blocks. The general trend for the encryption time is increased linearly with larger plaintext data size. However, an increase in the plaintext data size causes a decrease in encryption throughput. Typically, when the size of plaintext data is increased by 4, the encryption time rises by a factor of 4.33 and the throughput is reduced by a factor of 4.31.

An optimized AES CCMP, with interleaved CBC-MAC, was then implemented and the performance gain compared with the unoptmised version of AES CCMP. The enhanced AES CCMP is the combination of an optimized CBC MAC, while the Counter mode is unchanged. The Interleaved CBC (ICBC) in which multiple streams of CBC encryption are interleaved is motivated from the work by [K Gaj et al., 2000] which proposes the concept of parallel computation. The simulation results demonstrate that 2-Way Interleaved optimized AES CCMP exhibits lower encryption time and higher throughput compared to the unoptimized AES-CCMP implementation. Since encryption time is proportional to the energy consumed, we conclude that 2-Way Interleaved optimized AES-CCMP will preserve battery energy consumption for the same plaintext encryption task.

## VIII. Acknowledgments

The authors thank anonymous reviewers for their comments to improve the paper.

## IX. References

A. AZIZ AND N. IKRAM, "An FPGA- based AES-CCM Crypto Core for IEEE 802.11i Architecture", *International Journal of Network Security*, Vol5, No2, Sept 2007

A. SAMIAH, A. AZIZ AND N. IKRAM, "An Efficient Software Implementation of AES-CCM for IEEE 802.11i Wireless Standard", *31st Annual International Computer Software and Applications Conference* - Vol. 2- pp. 689-694, COMPSAC 2007

C.MUCCI, L.VANZOLINI, F.CAMPI, A. LODI, A. DELEDDA, M. TOMA AND R. GUERRIERI, "Implementation Of AES/Rijndael On A Dynamically Reconfigurable Architecture", *Design, Automation & Test in Europe Conference & Exhibition*, 2007

C.N.MATHUR AND K.P. SUBBALAKSHMI, "Energy efficient wireless encryption", IEEE *Global Telecommunications Conference,* 2006.

C.T.R. HAGER, S.F. MIDKIFF, J.-M PARK, T.L. MARTIN, "Performance and Energy Efficiency of Block Ciphers in Personal Digital Assistants", *In proceedings of the 3rd IEEE International Conference on Pervasive Computing and Communications (Percom 2005)*, IEEE Computer Society Press, 2005

Federal Information, Processing Standards Publication 197, "*Announcing the ADVANCED ENCRYPTION STANDARD (AES)*", November 26, 2001

H. CAM. S. OZDEMIR, D. MUTHUAVINASHIAPPAN, P. NAIR, "Energy efficient security protocol for wireless sensor networks", VTC 2003-Fall. 2003 IEEE 58th, *Vehicular Technology Conference*, 2003.

H. YANG, F. RICCIATO, S. LU, L. ZHANG, "Securing a Wireless World", Computer. Sci. Dept., Univ. of California, Los Angeles, CA, USA, *Proceedings of the IEEE, Feb. 2006 Volume*: 94, Issue: 2

J. GROßSCHÄDL, S. TILLICH, C. RECHBERGER, M. HOFMANN, AND M. MEDWED, "Energy evaluation of software implementations of block ciphers under memory constraints", *Design, Automation & Test in Europe Conference & Exhibition*, 2007.

J. NECHVATAL, E. BARKER, D. DODSON, M. DWORKIN, J. FOTI, E. ROBACK, "Status Report on the First Round of the Development of the Advanced Encryption Standard", *Journal of Research of the National Institute of Standards and Technology*, Volume 104, Number 5, September–October 1999

K. ATASU , L. BREVEGLIERI , M. MACCHETTI, "Efficient AES Implementations For ARM Based Platforms", *Proceedings of the 2004 ACM symposium on Applied computing*, 2004

K. GAJ, P. CHODOWIEC, "Hardware performance of the AES finalists - survey and analysis of results", *Technical Report, George Mason University*, Sep 2000, http://ece.gmu.edu/crypto/AES_survey.pdf

M DWORKIN, Recommendation for Block Cipher Modes of Operation: The CCM Mode for Authentication and Confidentiality, NIST Special Publication 800-38C, May 2004

N.R. POTLAPALLY, S. RAVI, A. RAGHUNATHAN, N.K.JHA, "Analyzing the energy consumption of security protocols, *Proceedings of 8th International Symposium on Low Power Electronics and Design*", ISLPED '03, ACM Press 2003

N.R. POTLAPALLY, S. RAVI, A. RAGHUNATHAN, N.K.JHA, "A study of the energy consumption characteristics of cryptographic algorithms and security protocols", *IEEE Transactions on Mobile Computing*, Vol 5, No 2, February 2006

P AGRAWAL, "Energy efficient protocols for wireless systems", *IEEE International Symposium on Personal, Indoor, and Mobile Radio Communications (PIMRC)*, Vol 2, Boston , USA, 1998

P. HAMALAINEN, M. HANNIKAINEN, T.D. HAMALAINEN, "Efficient Hardware Implementation of Security Processing For IEEE 802.15.4 Wireless Networks", *48th Midwest Symposium on Circuits and Systems*, 2005.

P. NI AND Z. LI, "Energy cost analysis of IPSec on handheld devices", Microprocessors and Microsystems, *Special Issue on Secure Computing Platform*, 2004.

P. PRASITHSANGAREE, P. KRISHNAMURTHY, "Analysis of energy consumption of RC4 & AES algorithms in wireless LANs", *Global Telecommunications Conference*, GLOBECOM IEEE, 2003

R. CHANDRAMOULI, S. BAPATLA, K. P. SUBBALAKSHMI, R. N. UMA, "Battery Power-aware Encryption", *ACM Journal Name*, Vol. V, No. N, February 2005.

M.R. DOOMUN, K.M.S SOYJAUDAH, "Adaptive IEEE 802.11i security for energy security optimization", *The Third Advanced International Conference on Telecommunications*, 2007. AICT 2007.

R. KARRI AND P. MISHRA, "Analysis of energy consumed by secure session negotiation protocols in wireless networks", *International Workshop on Power and Timing Modeling, Optimization and Simulation*, Torino, Italy, Sep 2003

W ROCHE, "The Advanced Encryption Standard, The Process, Its Strengths and Weaknesses", University of Colorado, Denver, *Spring 2006 Computer Security Class*, CSC 7002, Final Paper, May 6, 2006

Z. LI. R. XU, "Energy Impact of Secure Computation on a Handheld Device", *IEEE 5th International Workshop on Workload Characterization*, 2002

## X. X. Acronyms

| | |
|---|---|
| AAD | Additional Authentication Data |
| AES | Advanced Encryption Standard |
| ASIC | Application Specific Integrated Circuit |
| CCMP | Counter Mode (C) with Cipher Block Chaining (C) Message Authentication Code (M) Protocol (P) |
| CBC | Cipher Block Chaining |
| CPU | Central Processing Unit |
| DReAM | (D)ynamically (Re)configurable (A)rchitecture for future (M)obile communications applications |
| FIPS | Federal Information Processing Standard |
| FPGA | Field-Programmable Gate Array |
| GF | Galois Field |
| GHz | Gigahertz |
| HD Cipher | High Diffusion Cipher |
| ICBC | Interleaved Cipher Block Chaining |
| IEEE | Institute of Electrical and Electronics Engineers |
| IV | Initialization Vector |
| IPSec | Internet Protocol Security |
| LAN | Local Area Network |
| MAC | Message Authentication Code |
| MB | Megabyte(s) |
| MHz | Megahertz |
| MIC | message integrity code |
| NIST | National Institute of Standards and Technology |
| NOVSF | Non-blocking Orthogonal Variable Spreading Factor |
| PDA | Personal Digital Assistant |
| RAM | Random Access Memory |
| RC2 | Rivest Cipher 2 |
| RC4 | Rivest Cipher 4 |
| RC6 | Rivest Cipher 6 |
| RF | Radio Frequency |
| TCP/IP | Transmission Control Protocol/Internet Protocol |
| WLAN | Wireless Local Area Network |
| WTLS | Wireless Transport Layer Security |
| XTEA | eXtended Tiny Encryption Algorithm |